\documentclass[runningheads]{llncs}
\usepackage{epsfig,endnotes, graphicx}
\usepackage{amsmath}
\usepackage[dvipsnames]{xcolor}
\begin{document}

\newif\ifdraftmode
\draftmodetrue

\title{{\Large \bf  Intentional Forgetting}}

\author{Deborah Shands \and Carolyn Talcott} 

\authorrunning{D. Shands and C. Talcott} 
\titlerunning{Intentional Forgetting} 

\institute{SRI International}

\maketitle


\newcommand{\epsfigalt}[3]{ \begin{figure}[htbp] \begin{center} \leavevmode \epsffile{#1} \caption{#2\label{#3}} \end{center} \end{figure} }


\newcommand{\todo}[1]{{\color{red}\textbf{***TODO***} #1}}
\newcommand{\carolyn}[1]{{\color{red}\textbf{CAROLYN} #1}}
\newcommand{\deborah}[1]{{\color{red}\textbf{DEBORAH} #1}}

\newenvironment{TBD}{\begin{quote} {\bf To Be Done:} \em}{\end{quote}}   
\newcommand{\fixit}[1]{\mbox{$<$ \bf FIX IT: \it #1$>$}}
\ifdraftmode
\else
   \renewcommand{\todo}[1]{}
   \renewcommand\carolyn[1]{}
   \renewcommand\deborah[1]{}
\fi

\def\tbf{\texttt{tbf}~}
\def\prepare{\texttt{prepare}~}
\def\forget{\texttt{forget}~}
\long\def\omitthis#1{\relax}



\newenvironment{defn}{\begin{define} \rm}{\end{define}}
\newcommand{\eg}[1]{\mbox{\sl #1}}

\begin{abstract}

Many damaging cybersecurity attacks are enabled when an attacker can access residual sensitive information (e.g. cryptographic keys, personal identifiers) left behind from earlier computation. Attackers can sometimes use residual information to take control of a system, impersonate a user, or manipulate data.  Current approaches to addressing access to residual sensitive information aim to patch individual software or hardware vulnerabilities. While such patching approaches are necessary to mitigate sometimes serious security vulnerabilities in the near term, they cannot address the underlying issue: explicit requirements for adequately eliminating residual information and explicit representations of the erasure capabilities of systems are necessary to ensure that sensitive information is handled as expected.

This position paper introduces the concept of intentional forgetting and the capabilities that are needed to achieve it. Intentional forgetting enables software and hardware system designers at every level of abstraction to clearly specify and rigorously reason about the forgetting capabilities required of and provided by a system. We identify related work that may help to illuminate challenges or contribute to solutions and consider conceptual and engineering tradeoffs in implementations of forgetting capabilities. We discuss approaches to modeling intentional forgetting and then modeling the strength of a system's forgetting capability by its resistance to disclosing information to different types of detectors.  Research is needed in a variety of domains to advance the theory, specification techniques, system foundations, implementation tools, and methodologies for effective, practical forgetting. We highlight research challenges in several domains and encourage cross-disciplinary collaboration to one day create a robust theory and practice of intentional forgetting.  
\end{abstract}

\section{Introduction}
\label{intro}

Many damaging cybersecurity attacks  occur when the attacker accesses residual evidence of sensitive information (e.g. cryptographic keys, personal identifiers) left behind on a system from earlier computation. Access to such information can enable attackers to take control of a system, impersonate a user, or manipulate data. Often, that information persists unexpectedly: at one semantic layer, the information disappears (e.g., the variable goes out of scope, the file is deleted), but the layer below contains some remaining evidence of that information  (e.g., the last value of the variable still resides in memory, the bits representing the data from the file are still stored together on the drive).  At the higher level of abstraction, the system behaves as expected (e.g., the compiler indicates that the variable does not exist, the operating system indicates that the file does not exist), but at the lower level, a curious observer would likely find considerable evidence of the information.

Computer systems are designed to work this way for good reason: Multiple layers of abstraction enable the creation of logically complex behavior from the simplest of electronic hardware operations.  It seems irrational to demand that system implementations hide messy details and provide logically consistent abstractions except when we suddenly (and without warning) want the opposite -- high assurance that those messy details have very specific and previously unspecified properties. 

Current approaches to addressing access to residual sensitive information aim to patch individual software vulnerabilities or implement a software mitigation to fix an individual hardware vulnerability. Clever cybersecurity researchers have found alternative vulnerabilities \cite{2018-Lipp-Meltdown,2019-Kocher-Spectre} even after the fixes have been applied, challenging the system defenders to the next round of cat-and-mouse. While such patching approaches are necessary to mitigate sometimes serious security vulnerabilities in the near term, they cannot address the underlying issue:  without explicit requirements for adequately eliminating residual information, system designers must guess the capabilities architects of the next layer of abstraction might need.  

This position paper argues for an  \emph{intentional forgetting} capability.  Application software developers require an intentional forgetting capability in the operating system software.  System software designers require that capability in the hardware.  Integrating hardware to create a system requires specific intentional forgetting capabilities of the various hardware components, such as memories, processors, storage devices, or network hardware. Intentional forgetting capabilities and limitations can become discriminators for candidate implementations, just as storage capacity, computational speed, or message throughput are used today.  


Intentionality is essential to effective forgetting; systems must have the necessary forgetting capability, and systems layered on top of the implementing system must give advance notice of intent to use that capability.  Implementations of remembering provide an analog: for a computer program to remember a value throughout a computation, the programmer first declares a variable, providing an identifier and, often, a type, which indicates an intent to store and refer to the named data in the future.  When a runtime system executes the program in the context of an appropriate  operating system and computer hardware, some hardware memory is reserved.  As the program executes, it may store data in the memory that was reserved and named by the declared variable. The capability to remember the value of a variable is  offered to the programmer through features of a programming language and supporting system software and, finally, implemented in hardware.  The declaration of the intent to store a value leads the layers of implementation to \emph{instrument} the underlying system software and hardware to prepare to remember a value to be supplied later. Programming languages with implicit typing often compile first to an intermediate bytecode that enables the bytecode interpreter to identify and  reserve the necessary memory space.  Regardless of the approach used to reserve memory space for future use, programmers specify the intention to use memory to store data; the system software and hardware  instrument the system to make that use possible.

Intentionality is also required to enable concurrency control in distributed systems.  If concurrent processes share a resource (e.g., communicate via a shared variable), it is often necessary to ensure that only one process at a time accesses (reads or writes to) the resource at a time to prevent undesirably unpredictable system behavior.  Parallel programming languages offer a method for excluding access to the resource unless a process has first acquired some form of a lock (or mutex or semaphore).  Other processes must wait until the lock is released before they can access the resource.  The programmer must first declare and name a lock, which processes can later request to acquire.  While different operating systems provide different implementations of locking, all require advance system instrumentation to enable future use by executing processes.

Intentionality in forgetting is displayed by systems that implement transactions.  If Alice wants to transfer \$10 from her bank account to Bob's account at a different bank, a number of intermediate actions must take place, including the reduction of the value of Alice's account and the increase in value of Bob's account.  At the end of the day, regardless of computer, software, or network failures, both banks must account for the disposition of the \$10 in the same way; it must appear in either Alice's account or Bob's account and all of the intermediate steps of the money transfer are ``forgotten'' from Alice's and Bob's points of view.  Enabling the forgetting required for transactional semantics (``atomicity'') requires significant software instrumentation in the form of multi-phase commit protocols and their implementations.  While atomicity is achieved at the application or user level, the banking system logs do not forget.  All of the records of intermediate activity are logged to enable a future audit; intentional forgetting is achieved within one level of abstraction, but is explicitly avoided at the implementation layer below.  

The remainder of this paper discusses the capabilities that are needed to enable intentional forgetting.  We identify work in some research domains that may be applicable and call attention to some of the challenges that must be addressed to produce viable solutions. Section~\ref{specifying} discusses a two phase-based specification of forgetting  and compares with other systems that operate in two phases: an application first declares an intention to take future action and, later, calls upon the system to act. Forgetting capabilities may be easier to implement in some systems if specifications enable system instrumentation in a phase prior to forgetting actions. Section~\ref{instrumenting} highlights some current approaches to instrumenting systems to track information.  The conceptual and engineering tradeoffs in instrumentation approaches may help to guide consideration of tradeoffs that are likely to arise in implementations of forgetting. Section~\ref{modeling} focuses on the modeling of intentional forgetting at the abstraction and implementation layers and the coordination and composition of forgetting capabilities. Detectors, discussed in Section~\ref{detectors}, model the information  recovery capability of different  adversaries.  A determined adversary may present a much more capable detector (e.g., well-resourced, clever) than a casual observer, so presenting a system's forgetting capability in terms of its resistance to different types of detectors provides a means of expressing the strength of forgetting. Section~\ref{relatedwork} touches on areas of related work, and Section~\ref{opportunities} outlines research opportunities, ranging from theory and modeling to system performance.  Section~\ref{conclusion} concludes with a brief summary.

\section{Specifying intentional forgetting}
\label{specifying}

The examples of programming, concurrency control, and distributed bank transactions in Section~\ref{intro} contain a \emph{prepare}~phase, followed by an \emph{action} phase. The abstraction layer calls some form of \prepare\ command, which causes the implementation layer to begin instrumenting the system to be ready to later respond to an appropriate command for action. The action command of interest to us is the \forget command. Although a universal language or method of specifying intentional forgetting is not likely possible, we expect the prepare/forget phases and commands to follow a common pattern across a variety of abstraction and implementation layers. 

Consistent with their usual behavior, the abstraction and implementation layers dictate the syntax and approach to  indicating an item that  is \emph{to be forgotten} (or \tbf) and then, later, to request that the forgetting take place.  For example, when the abstraction layer is a high-level programming language and the implementation layer is an operating system, the programming language might include the reserved word {\bf TBF} for designating a variable {\tt x} as \tbf at the time the variable is declared.  Later, the program makes a system call {\tt forget(x)} to cause the value of {\tt x} to be forgotten. 

Consider rollbacks of distributed transactions, for example.  After nodes supporting a distributed transaction are asked to {\tt prepare}, each node executes instructions to allocate local resources, record redo logs, and set locks.  Later, each node receives either a {\tt commit} instruction, which causes the node to make permanent its local portion of the transaction, or a {\tt rollback} instruction, which causes the node to return to the same local state as before the \prepare  instruction was received. A node uses the information recorded in its local redo log to correctly reset any local values to their earlier state.  Rollbacks are often used to address failures (often network failures) that happen while a transaction is executing to ensure that the system as a whole is not left in an inconsistent state.  Rollback is an form of forgetting that tidies the abstraction layer but does not clean up residual evidence of an attempted transaction at the implementation layer. System logs will show the intermediate steps that were attempted to execute the transaction and can be used to diagnose problems in the system.  Because rollback does not tidy the implementation layer, it does not address the challenge of cross-layer forgetting that motivates this study.  However, the prepare/rollback pattern that distributed systems programmers use to ensure correctness in transactional semantics is similar to the prepare/forget pattern that is needed to provide cybersecurity assurance of forgetting. 

If the implementation layer is a hardware device, then it may offer a {\tt forget} command via a device driver installed in an operating system.  The language of the \forget command would be similar to that for other operations of that hardware device. 

In each of these examples, the \prepare  and \forget commands are made available for explicit use by the abstraction layer and are implemented by the layer below. The nature of the specification fits the particular abstraction and implementation layers involved.

\section{Instrumenting systems for intentional forgetting}
\label{instrumenting}

To thoroughly forget an information item on demand,  the system must be able to identify all the item's  implementation layer elements of the item and  take action to remove  them. A \prepare command initiates a prepare phase during which the implementation layer instruments the system.  After the prepare phase, the implementation continues to manage the instrumentation to enable later forgetting.  
Instrumentation takes many forms depending on the system and the type of information to be forgotten. Computer systems have typically used one of two classes of instrumentation techniques to enable information to be deleted. 

The first approach involves marking an item so it can be traced as it is manipulated within the system.  Examples of systems that mark include 1) garbage collection \cite{1960-McCarthy-GarbageColl} and Automatic Reference Counting (ARC), which is implemented by the LLVM compiler for Objective C \cite{Objective-C-ARC}; 2) page replacement algorithms for virtual memory managers \cite{2017-Lustig-ArchOSSupportVM} that mark a modified memory page with a modified or dirty bit to note the need to write that page to stable storage before swapping out the page and reclaiming the memory space; 3) dynamic  taint tracking \cite{2010-Schwartz-DynamicTaint}, which can trace the propagation of data objects during program execution to, for example, identify the impact of un-validated input data; and 4) static analysis tools \cite{2000-Engler-CompilerExtensions,1999-Foster-TypeQualifiers,2005-Halfond-AMNESIA,1999-Myers-JFlow} that use data flow  techniques to track the use of data within a program.  All of these marking techniques are fine-grained, enabling systems to track and address individual data items or their implementation-level representations.  For example, a garbage collector can determine that a specific integer variable is no longer in use and reclaim its memory space for other uses.

A second approach is coarse-grained, clustering together items and treating them similarly at the implementation layer; all of the items in the cluster are either remembered or deleted.  For example, magnetic drives can be degaussed to delete their entire contents.  Flash-based solid state drives (SSDs) require different techniques \cite{2011-Wei-ErasingFlashDrives} to ensure that data is removed.  This coarse-grained approach is currently used by encrypted memories, firmware and self-encrypting drives to reliably delete a large amount of data by forgetting only a single encryption key. 

Both the fine-grained marking approaches and the coarse-grained approaches  require  ongoing system instrument management.  Instrument management consumes  resources, which leads to various engineering tradeoffs in system design. Specialized, resource-constrained systems often forego most such techniques, due to resource constraints. Experience with a variety of system instrumentation and management approaches indicates that a forgetting capability is likely to be expensive to implement, so research into techniques for optimizing forgetting instrumentation is of great practical importance.


\section{Modeling intentional forgetting}
\label{modeling}

Currently, application developers and  users do not have a basis for forming realistic expectations about the residual system information that might be found if someone were to look for it. Some systems offer weaker qualities of forgetting, which yield information to a curious amateur, while other systems withhold the information from all but the most determined professional sleuth.  In short, application developers and users need a clear description of the forgetting capabilities of the system.  Capabilities of interest include the granularity of information that can be forgotten (i.e., can the system be instructed to forget just an integer or does it forget much coarser ``chunks'' of information?) and  the strength or quality of the forgetting (i.e.,  how hard is it to recover information that the system is instructed to forget?)

To provide rigorous specification of forgetting capabilities to support precise communication, we recommend the development of formal models to describe the abstraction layer, the implementation layer, the composition of forgetting capabilities as systems are constructed from components, and the actors that seek residual information. For concreteness in this section, we focus on an abstraction layer where a programmer writes software and the corresponding  implementation layer that provides computation, storage, networking and other services that will be used to implement the application logic.

\omitthis{Andy asks for diagram}

At the abstraction layer visible to the programmer, a possible approach is to model behavior of
a system design as an abstract machine (an automaton) that specifies changes in state and interactions with the external world.
An implementation of an abstract machine on a more concrete machine may have intermediate states using information not visible at the abstract level. 
The language runtime system, operating system, CPU, and other hardware that interpret the implementation machine may expose even more information. For example, speculative execution or manipulation of a memory management system may expose timing information or leave traces of sensitive data.
Stuttering (bi)simulations can show that a lower layer
(implementation) is correct with respect to the upper layer
(specification).\footnote{A bisimulation is a 1-1 correspondence between states
of executions at the two layers. In a stuttering bisimulation, at one layer, usually
the lower, states may be skipped.}
Can such relations be augmented to show that the lower layer
adequately implements forgetting? 

Here are some questions to consider when addressing intentional forgetting.
\begin{itemize}
\item
How can the programmer or designer annotate data that should be forgotten after the program or process is done with it?
The annotation must be propagated  along the execution and to
the lower levels. The annotations should be able to express
different levels of sensitivity which would then result in
instrumentation for different levels of forgetting. 
The programmer may also wish to express a minimal level of forgetting that
is acceptable, to allow some flexibility.
\item 
What program structures might be better able to support forgetting (e.g. scoping mechanisms or data abstractions)?
\item 
What additional information (beyond the instruction API) does a compiler need to
know about its target platform (operating system and hardware) to support
instrumentation of the machine-level code to achieve sufficient forgetting?
\item 
A model of accessibility and information extraction ability/cost (attack model)
is needed to say when something has been forgotten.
For example, all the memory may be accessible but how does one
know which blocks represent information of interest.  How much work is
required to extract usable information?
\end{itemize}

\omitthis{Need analysis tools to catch programmer mistakes.}

To discuss forgetting requirements and challenges in more detail,
the following section introduces some mathematical notation for
describing processes at two layers (abstraction and implementation)
and the relationships between them. Section~\ref{detectors},
discusses detectors as measures of forgetting and attack strength.

\subsection{Modeling the abstraction and implementation layers}

The abstraction layer has a state transition
automaton with states $S = q,D$ ($q$ for abstract program
state, $D$ for data state) and actions $a$, which include
input, output, and application of data transformations. 
For purpose of illustration, 
$D$ is assumed to be a data dictionary, i.e., a mapping from
names to data structures and the named entities 
will be the unit of forgetting. An execution trace
for such an automaton has the form 
$$ S_0 \xrightarrow{{a_1}} \dots \xrightarrow{{a_k}} S_k$$ 
where
the action $a_{i+1}$ is determined by the automaton
transition system, the state $S_i$, and, possibly, external input.

To prepare for forgetting, the programmer must specify
\begin{itemize}
\item
   \tbf data initially present in $S_0$ by specifying
which names bind \tbf data 
\item
 the effect of actions, whether they introduce new \tbf data (such as reading or generating a secret), and how a primitive operation or function call propagates \tbf status 
\item
conditions that trigger forgetting, such as program exit
\end{itemize}
\noindent
The \tbf status of a named data item may include a desired
level of forgetting. The level relates to the computational
cost of forgetting, and to the difficulty of detecting.
Making these relations explicit is important for making
both conceptual and engineering tradeoffs.

\omitthis{Need diagram relating S,M, i, i_j}

The \tbf annotations of the initial state and actions
allow the \tbf annotations to be automatically propagated to
newly created and transformed data. For an execution as
above, $St_j$ for $1 \le j \le k$ denotes the result, $St_0$, of
annotating $S_0$ and propagating annotations according to
annotation of actions.

The implementation layer contains a state transition
machine with states $M_i$ that include elements $(pc, regs,
mem)$ where $pc$ is the program counter, $regs$ is the register set,
and $mem$ includes cache levels and any temporary
storage used in execution. Transitions are labeled by
machine actions $x$ 
(processing instructions, input-output actions, \dots). 
$$M_0 \xrightarrow{x_1} \dots \xrightarrow{x_l} M_l$$
In general, $k < l$ as a program level action may be implemented
by multiple machine-level instructions.

Assume that $\sim$ is a (stuttering bi) simulation relation between abstract states $S_i$ and a subset of implementation states $M_{i_j}$
showing the correctness of the machine model (and compilation).
Thus if $M_0$ is  the compilation of $S_0$ then $S_0 \sim M_0$.
And if
$$S_0 \xrightarrow{{a_1}} \dots \xrightarrow{{a_k}} S_k$$ 
then there is a subsequence of indices $i_j$ for $1 \le j \le k$ 
such that  $S_j \sim M_{i_j}$  for $1 \le j \le k$.
$$M_0 \xrightarrow{\overline{{x_1}}} \dots  
      \xrightarrow{\overline{{x_k}}} M_{i_k}$$
where $\xrightarrow{\overline{{x_j}}}$ indicates a sequence of machine
instructions and intermediate states,
and $i_j$ is an increasing sequence of indices picking out the
simulation points.  Thus $i_j >= j$ and $k <= i_k <= l$.
For example, the machine state corresponding to abstract state $S_1$
is $M_3$, i.e. two machine states are intermediate instructions,
thus for $j=1$, $i_1 = 3$.

Relevant properties of $S_i$ are preserved by the $\sim$ relation as
corresponding properties of $M_{j_i}$ which capture the sense in
which the machine is a correct implementation. Properties could
include correctness of data representations, data invariants, or
axiomatic theories to be satisfied by both levels.

The annotation of abstract actions needs to be compiled to
instrumentation of implementing machine instructions that
determine how \tbf annotation is introduced and
propagated. Points where a \tbf class should be forgotten
should be compiled to suitable forgetting instructions.

Suppose memory blocks of some convenient size can be
\emph{colored} with a finite set of colors. Then
 memory holding \tbf data could be colored according to 
 forgetting class, such that data of
 the same color can be forgotten together, at the same level. For example, let
$Mc_0$ be $M_0$ with the image of \tbf data in $St_0$
 (i.e., the first state in the annotated execution trace) colored according to the forgetting class to which the named data
is assigned.
The idea is to lift the simulation relation $\sim$ to a
relation  on annotated program level states and
colored machine-level states.
First we require $St_0 \sim Mc_0$.
Next we need to propagate colors using annotated machine
instructions and check that at simulation points $S_j \sim
M_{i_j}$, we have $St_j \sim Mc_{i_j}$.
In particular, in addition to preserving functional properties
(that should not be affected by annotation or instrumentation),
the image of \tbf data in $St_j$ is suitably colored in 
$Mc_j$. The image of a data element needs to 
include its total footprint in memory (including cache) and registers (that 
have not been overwritten by other data).
Thus,
\begin{itemize}
\item
if $S_j \xrightarrow{{a_{j+1}}}  S_{j+1}$ 
adds a \tbf structure then its image in $M_{j+1}$ should
be suitably colored
 
\item
if $S_j \xrightarrow{{a_{j+1}}}  S_{j+1}$  forgets'a level then the image of data of the corresponding color in $M_{j+1}$ should be  not detectable (i.e., no information about the \tbf data can be detected (with the given effort)
 
\item
if $S_j \xrightarrow{{a_{j+1}}}  S_{j+1}$  reads  \tbf data 
from an external source, then the corresponding read
instructions $M_{i_j} \xrightarrow{\bar{x}} M_{i_{j+1}}$ should color every storage location including temporary locations such as buffers, registers, or cache, in which the \tbf data resides
 
\item
dually, if \tbf data is transmitted to an external target (external machine, database, file) actions beyond the local interface can not be
controlled by local forgetting, but every storage location in which the data resides (even temporarily) must be colored and managed by the local forgetting machinery.  
\end{itemize}

\omitthis{
Andy says need IFML extension to HTML (Intentional Forgetting ML) recipient 
of data must be trusted.
}

\begin{figure}[ht]
\centering
\includegraphics[width=0.9\textwidth]{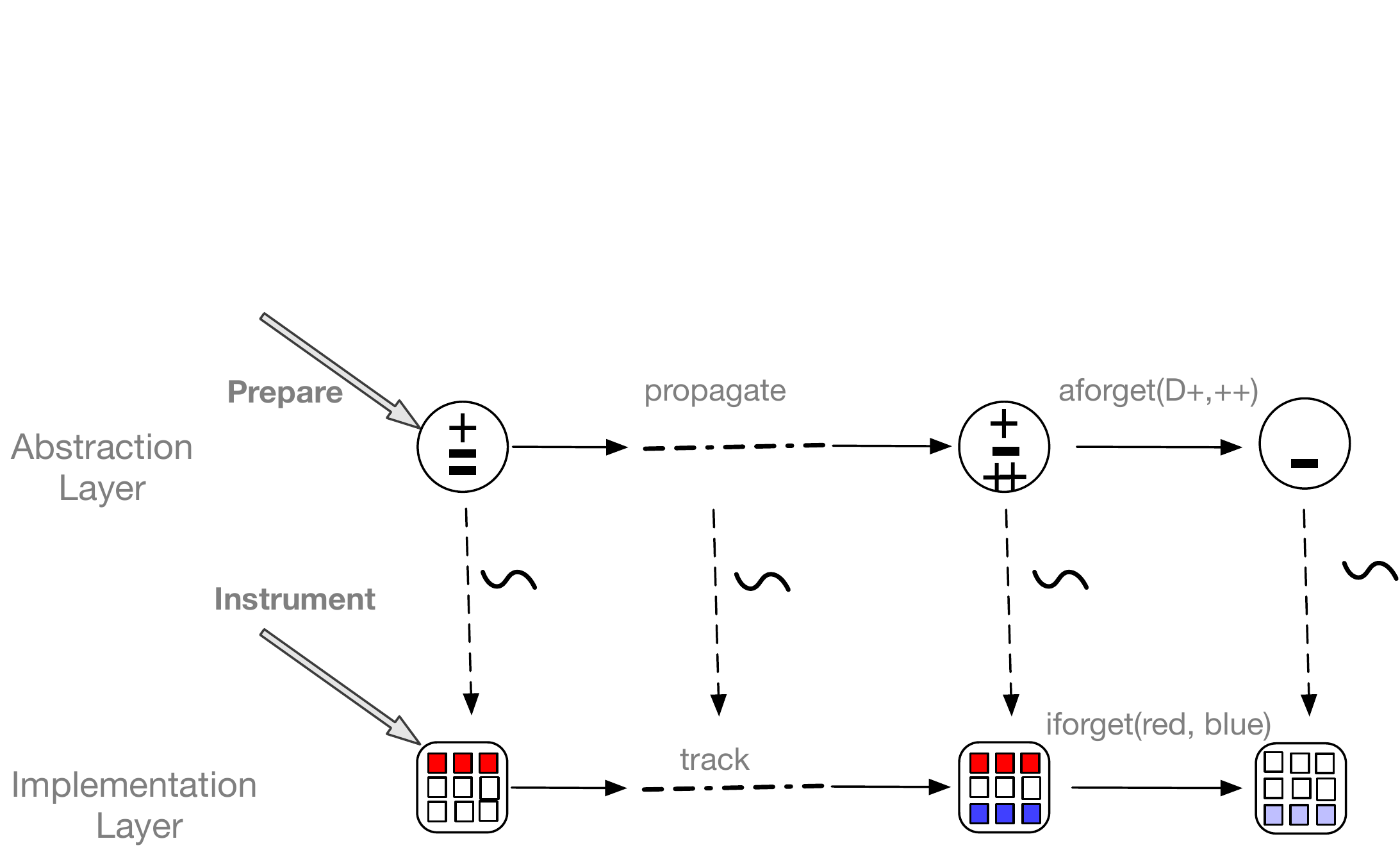}
\caption{{\bf Simulation of a forgetting aware abstract machine.} 
At the abstraction layer, circles represent abstract
states and \texttt{+,++} 
\tbf data requiring high and low strengths of forgetting.
The corresponding implementation data is indicated by red and blue  within rectangles representing the implementation state. }
\label{ffig}
\end{figure}

Figure \ref{ffig} illustrates these ideas with two forgetting capabilities of different strengths: level blue, which deletes a file by removing its entry in the file index; and level red, which deletes the file using a cryptographic erasure technique.

\noindent  
Figure~\ref{ffig} shows that the programmer's preparation
marking data at the abstract level is reflected by the 
implementation-level instrumentation (coloring). 
The vertical arrows labelled by the simulation sign
show related abstract and implementation states.
The propagation of annotations at the abstract level
is reflected by the tracking done by the instrumentation.
When the abstract level specifies forget at the \texttt{+} level, 
 then the implementation executes
\texttt{iforget(red)} (the file is unrecoverable), and when 
\texttt{aforget(++)} is specified, the implementation executes
\texttt{iforget(blue)}.   The \texttt{grey} indicates that the
file is not erased but is not accessible via the file index.

Some things are outside the direct control of machine-level
instructions. We assume that that the registers and local data of an
executing process is suitably protected from access by external
processes. Other issues must be considered.

\begin{itemize}
\item
One example is virtual memory.  When a page is brought in, presumably
its remote copy is freed and any colored blocks must be forgotten.

\item
If state is check-pointed for exploratory computation, any colored blocks
copied must be forgotten if the computation fails and the original state is restored.
 
\item
A process
running in a multiprocessing environment is subject to 
being interrupted or paused while another process runs.
The forgetting mechanism must account for \tbf
data contained in saved state and make sure no
\tbf data is left exposed when state is restored.
In addition \tbf data in the unsaved process state should 
be forgotten  during execution of the interrupting process;
i.e., it should not be accessible to the foreign process.
\end{itemize}

\subsection{Coordinating and composing forgetting capabilities}

The preceding section focused on basic sequential processes which
are already complex and challenging. Intentional forgetting gets
even more complicated when an application or system is composed of a
(possibly distributed) network of interacting processes or when the
hardware supporting the implementation is considered to be made up
of interconnected components (e.g., CPU, GPU, RAM, secondary
storage, network interfaces), each with different capabilities for
remembering and forgetting.

In the case of interacting processes \tbf data may flow between
processes. Thus forgetting instrumentation must support propagation
of \tbf annotation and forgetting instructions across process
boundaries. Also, communication mechanisms, such as shared memory,
buffers, pipes, buses, and sockets, must provide support for forgetting.
The overall achievable level of forgetting will depend on which data
flows across the interfaces, the level of forgetting
that can be achieved by each process, and the forgetting achievable
by the communication mechanisms. 
One question is whether the overall
forgetting level is the minimum of the component and communication
forgetting levels or if the minimal level can be improved. Using dataflow analysis should make it possible to identify \tbf data that remains local to a
process and can be forgotten locally. For example, secure
communication protocols must ensure that keys and other sensitive
material is not leaked in transit, when sessions end, if
negotiations are aborted, or if power fails. Protocol designs and
implementations would benefit from systematic, semantically defined
annotations to automate much of the needed protection and provide
input for verification tools.

In addition to information flow, 
information may be split across entities so that no one
entity has a sufficient share to be considered useful.
What constitutes forgetting in this case? Is it possible to make it
sufficiently difficult to obtain enough shares to
reconstruct useful information without requiring that all shares be forgotten?

\omitthis{
Andy says --hard to automate, easy to get wrong manually
}

It maybe that actions must be transactional. In this
case, processes must either succeed or forget that
they tried, including forgetting all changes. Transaction
protocols such as two-phase commit are examples of mechanisms to prepare for forgetting. Here, what is
forgotten is  a partially completed action,
which, if not forgotten, might leave sensitive information
behind and leave the world in an
inconsistent and/or undesired state. In addition to forgetting
(undoing) partially complete actions if the actions
involved sensitive data, additional instrumentation may be
needed to make sure this data is not accessible to
unauthorized entities.

To address forgetting in cases involving composing hardware components we must consider
\begin{itemize}
\item
 what \tbf data crosses which interfaces?
\item
 is data no longer \tbf after crossed a boundary, or  does it possibly require stronger forgetting?
 \item
 is some form of forgetting contract needed?  
\item
 where is the \tbf data propagated and does it move (i.e, its original location should forget the data) or replicate (i.e., the data is stored at multiple locations)? 
\item
the data must be suitably annotated so that the external
 entity knows how to protect it 
\item
 will exported \tbf data cross back to its source or cross another interface?
\end{itemize}

Flow of \tbf data into one or a bank of GPUs
is a particularly interesting challenge.  
What is the attack model?  What  possible forgetting
mechanisms are compatible with efficiency requirements?
Another challenge is storage systems that use striping
and replication across multiple partitions for fault tolerance.

When should \tbf data be allowed to cross an interface, for
example, from local memory to a database or across a
network boundary? This is a matter of higher level policy
and mechanisms to protect information in transit. It
requires  communicating entities to agreed on languages for specifying forgetting policies and supporting protocols and interfaces. What is relevant
for forgetting instrumentation and enforcement is making
sure that system interfaces are considered, and that lower level
mechanisms that cause information to cross interfaces are
exposed at higher levels to allow appropriate protections
to be put in place.

\section{Detectors as a measure of forgetting capability}
\label{detectors}

The concept of \emph{detector} is introduced as a measure of
information recovery capability. A detector observes and/or probes
implementation layer resources (e.g., machine state, register
values) for bits and changes that reveal information (keys, indications
of code executed, \dots).

\omitthis{better examples}

Detectors have several roles:
\begin{itemize}
\item characterizing forgetting by resistance to detection and/or recovery
\item measuring  the ``strength'' of  forgetting
\item modeling attack capability 
\item  formalizing forgetting guarantees
\end{itemize}

As a forgetting measure the detector provides a way of
characterizing how much data and what data or actions should be
forgotten to prevent detection, reduce the
confidence of an attacker, or make it prohibitively expensive.

As an attack model, a detector can be used to characterize
the ability of an attacker, including what information can
be extracted, the potential utility of the information and
how much effort the attack requires.
It can characterize and quantify an attacker's
ability to see residual information and  to distinguish between machine states, for example, before and after forgetting.

As a way to formalize a guarantee, 
a detector can be used to specify what a component can forget.
For example, a memory manufacturer could specify that it's product 
supports \texttt{d54} level forgetting.

It is not enough for a detector to observe bits, it must
interpret the bits to recognize them as providing
desired, useful information, such as an encryption key, a password,
or the highest bid on an item. For this
purpose, the detector needs some knowledge, such as
\begin{itemize}
\item
general knowledge about the size and shape of the bits it needs
\item
specific knowledge about the application code, the data structures
the code uses, what the code creates, reads, and stores
\item
knowlege about the machine, such as register and stack layout and
usage, memory management procedures and policies, etc.
\item
knowledge of standard access patterns that provide clues to the
operations being carried out.
\end{itemize}

In addition to what a detector can detect, given certain knowledge,
a detector is characterized by the amount of effort required to
succeed. Measures that are important for forgetting include the
efforts to detect, use, and forget information. Work measures
include time (how long detection takes), CPU cycles, memory, and
possibly access to other resources. The quality of the detection and
the detector's confidence in what was detected is likely to be
related to the detection effort and to the available information.

For example, an attacker that needs to decrypt cipher text might use a detector to find the corresponding decryption key stored in memory.  One detector could use a brute force approach to identify all bit sequences in memory that are plausible candidate keys.  A second detector could, instead, focus on regions of memory that are used by system services or applications that are known to perform encryption.  By focusing on limited regions of memory that have a higher likelihood of containing a cryptographic key, the second detector reduces the time and level of effort to produce candidate keys.

Another factor to consider is time of detection
\begin{itemize}
\item 
a detector that runs in parallel with an application, and can observe
changes in shared resources concurrently or when the application 
is paused or interrupted.
\item 
a detector that runs when application terminates and releases its resources
\item 
a detector that observes execution patterns by electrical signals
\end{itemize}
\noindent
Also, the time of detection relative to data generation
is important.  If the observation is too early, the relevant information is not
available, and if it is too late, the information could be diluted
by noise due to ongoing independent activities.
This implies considering a measure of skill on
the part of the user of a detector, or  additional
intelligent capabilities of the detector itself.

In  Figure~\ref{ffig}, the \texttt{++} (blue) detector
could  be a file recovery application that restores the ``deleted'' file's index entry to the file system's index so that it can be accessed as usual. 
A \texttt{+} (red) detector could be a sophisticated tool that 
that watches for user access to financial or medical information
and can access residual data stored in memory by a browser.
Forgetting techniques to thwart such a  \texttt{+} (red) detector are much more complex and expensive to implement than those needed to prevent the recovery of deleted files.

\paragraph{A Detector Algebra}
New detectors can be formed by combining existing detectors.
In some cases, one detector may provide knowledge for a
second detector to use, or, two detectors can use
independent methods to detect the same thing, to
increase confidence. In a distributed context 
multiple detectors observing at distinct locations
may be more powerful than a single detector.

Realizing the potential of detectors requires the development of a
theory of detectors, their capabilities, required resources, cost,
complexity, and composition. The theory can
incorporate and build on information theory, complexity theory, and
logical inference tools. In contrast to the formal analysis of
relations between abstraction layers where the set of behaviors is
bounded, detectors are trying to approximate what we do not know and
must expose.

A theory of detectors  must 
\begin{itemize}
\item
characterize the different kinds of information
that can be detected, 
\item
measure  information quality, including confidence,
probabilistic, and information theoretic measures
\item
characterize the knowledge and other resources needed to detect and,
dually, to forget
\item
provide measures of the work (hardness of the detection problem) required to
detect its target information, and, dually, the work needed to forget
\item provide mechanisms for combining and composing
detectors and for determining the properties of the resulting
detectors in terms of the properties of the components
\item
provide a means to measure
the effectiveness of different kinds and levels of forgetting
\end{itemize}

A key aspect of the theory will be a partial order on detectors,
$d_0 <  d_1$ (read $d_1$ is stronger than $d_0$). 
If $d_0$ can detect information $I$, and $d_0 <  d_1$
then  $d_1$ also detects $I$.
Dually, if instrumented code forgets to level $d_1$ 
and $d_1$ is stronger than $d_0$ then 
the instrumented code is secure
against an attacker of strength $d_0$.

The theory of detectors should not only provide a sound
and rich mathematical foundation for reasoning about
forgetting, it should also provide a practical foundation
for specifying \tbf information, for correctly
instrumenting code, showing that manufacturer's
forgetting claims are valid, and for showing that
instrumented code makes correct use of forgetting
capabilities.

\omitthis{Andy asks for example detector description.  Not really in scope?}

\section{Related work}
\label{relatedwork}

While our motivation for intentional forgetting is rooted in cybersecurity and privacy concerns, the need for the assurance of system correctness and reliability have motivated many computer science techniques such as those for ensuring modularity, scope of naming, synchronization of concurrent processes, and virtualization mechanisms. All such techniques provide tools to help people assert fine-grained control over the execution details of a system.  In some cases, the techniques enable automated tracking of items throughout a system execution so those items can be examined or otherwise managed when necessary.  In other cases, they enable aggregation of items to ensure that they are managed similarly.  Many of these techniques may inform the development of intentional forgetting capabilities; some may be directly adaptable to forgetting, while the fundamental insufficiencies of others may prove instructive.  The following sections highlight  a few of the many areas  that seem likely to inform efforts toward intentional forgetting.

\paragraph{Programming language semantics} A variety of programming language constructs have some relevance to intentional forgetting.  For example, many programming languages define the lexical scope, or region of a program, in which a name binding to a program object is valid \cite{2015-Scott-ProgLangPragmatics}. Outside that scope, the binding is forgotten (i.e., the name is forgotten).  Some approaches to lexical scope may prove helpful in identifying items that are \tbf outside a logical region of a program.  

 \paragraph{Language processing, execution environments, and tools for programmers} When intentional forgetting is applied at the level of application software, instrumentation techniques that assist in program execution may be relevant to intentional forgetting.  For example, garbage collection \cite{1960-McCarthy-GarbageColl} and ARC automate the process of software memory management. Both techniques instrument application code as it is running to reclaim system resources that are no longer in use.  In  \cite{2020-Cohn-Gordon-Facebook-deletion}, Cohn-Gordon and other authors from FaceBook discuss the challenge of robust deletion in online social networks and describe their DelF deletion framework that enables developers to annotate data type definitions and use the framework to automatically map deletion actions into asynchronous, reliable and temporarily reversible operations on backing data stores. All of these techniques track data items for abstraction layer tidying and may prove instructive for cross-layer forgetting.
 
\paragraph{Synchronization techniques} Forgetting requires control over the data associated with the state of the implementation layer machine. Many techniques for synchronizing processes in distributed computing or ensuring the state of a cyber-physical system address obligations for aligning states between two different types of executing systems. Such alignment is a critical element of cross-layer forgetting, as implementation layer data must be removed exactly (i.e., all of the implementation layer data that matches the abstraction layer item must be removed, but no more).

\paragraph{Current erasure methods} Although intentional forgetting has not yet been rigorously treated, a variety of techniques for erasure or data destruction \cite{2013-Reardon-SOK-DataDeletion} have emerged to partially address the need.  For example, self-encrypting memory or hard drives use encryption techniques to prevent detectors without a cryptographic key from reading stored data.  Many current techniques are likely to be viable implementations of intentional forgetting.  Paired with methods for rigorously specifying their capabilities (e.g., in terms of resistance to specific detectors), even weak implementations of forgetting may be acceptable in contexts where only weak detectors are present.  

\paragraph{Modeling techniques}

Foundations for modeling include computational models of
processes and programs at different levels of abstraction,
representation of properties, and relating different levels
and models.

The use of various automata (timed, IO, Mealy, Constraint, networked) and state transition systems is well known~\cite{automata}. Rewrite theories \cite{unified-tcs,maude-book}
provide a rich formalism for describing states at many levels of
detail, as well as representing transitions as rewrite rules. Such
models can be directly executed. In all of these cases a trace
semantics provides the basis for reasoning about execution properties.
Properties as sets of traces, expressed in various logics or as
mathematical formulae are the basis for reasoning about individual
runs. These can be mechanically checked using tools for reachability
analysis and model-checkers \cite{checking}. Hyperproperties
\cite{clarkson-schneider-2010jcs}, sets of traces, are
important to compare executions. They allow one to express
observational equivalence, information flow, and contracts.

In \cite{2020-Dullien-WeirdMachines}, a simulation relation is used to relate
an automata model and a simple machine model of a password storage application
to show that the machine can not be tricked into revealing a password.
In \cite{Spectre}, a microarchitectural model with execution semantics
is used to model information that can be extracted from normally
hidden microarchitecture structure and operations.

In \cite{pals}, a bisimulation relation is used to show
that under suitable conditions, a synchronous model of a
real time system satisfies the same properties as an
asynchronous, distributed model.

Observational equivalence is a notion of equivalence of programs, processes, or systems
based on an ability to observe.  What is observed could be
output, interaction sequences, or detectibles.  
This capability has been well developed for concurrent and distributed programs~\cite{actor-paper} and
for security protocols with and without timing considerations~\cite{cortier09csf,CC-post15,gazeau17esorics,nigam-talcott-urquiza-19cathyfest}.  

\section{Research opportunities}
\label{opportunities}

Intentional forgetting is essential to enable systems to meet the cybersecurity and privacy requirements of users.  It is a challenging capability for systems to provide and for software and hardware developers to use judiciously. Research is needed in a variety of different domains to advance the theory, specification techniques, system foundations, implementation tools, and methodologies for effective, practical forgetting. In particular, we see research challenges and opportunities in the following areas:

\paragraph{Theory, modeling, and formal methods}  Section~\ref{modeling} outlines several of the issues that arise in the formal modeling of forgetting across layers of abstraction.  It is always a challenge to capture with appropriate fidelity the relationship between an abstraction layer and an implementation layer, but forgetting requires consideration of model fidelity from the perspective of a detector (as discussed in Section~\ref{detectors}).  Producing models of detectors that parallel real-world use cases will be essential to establishing a useful theory of forgetting.  Recovery of data from accidental system damage (e.g., failures of hard drives and other components) is a very common use case, so modeling detectors that represent various data recovery paths are as important as modeling detectors that represent cyber attackers.  
Underlying models of detectors will be abstract machines and models of machine
architectures and their representation of information.
Three key challenges include 1) finding levels
of detail that expose what is important to consider from
an information detection perspective that are feasible to
reason about; 2) developing simulation relations that capture both data and information content (these will be essential for verifying forgetting claims); 3) defining abstraction relations
that allow us to prove properties of machine level
execution by reasoning at a more abstract program level
of execution (this will be important to make (semi-) automatic verification of forgetting properties feasible).

\paragraph{Methods of specification}  It is essential to specify not only what is \tbf but also the forgetting capabilities, expressed in terms of resistance to detection, of the implementation layer. However, code developed for one environment may later execute in a very different environment and may be combined with other components to create new systems.  Specification methods must support new compositions (e.g., system integration processes) and new environments (e.g., shifting from on-premises execution to a cloud-hosted environment)  to enable access to forgetting functionality and enable the assurance of the forgetting capabilities of the composed system.   Assurance might take the form of formal arguments, such as mathematical proofs, statistical arguments, or information flow arguments. System architecture  and physical properties also could contribute to assurance arguments.  
Research into methods for specifying intent to forget and the forgetting capabilities of a system, along with forgetting-related constraints on composition or deployment environments will be necessary to build a language for interested parties to communicate precisely about forgetting. An important component of this research
will be identifying conceptual abstractions at the
specification, program, and implementation levels that
can be successfully used by non-experts in designing
and developing systems.


\paragraph{System foundations} Important conceptual properties often require foundational support from the system. Distributed and other concurrent systems require synchronization primitives that might be implemented via shared memories, locks based on interrupt signals, or synchronization protocols. Cryptographic software relies on a system-provided source of randomness (or often, pseudo-randomness). Virtualization of every kind relies on a suite of system functions to record, map, schedule, copy, and signal. On what hardware and software system foundations will forgetting concepts rely? What new program abstractions will help ensure correct forgetting? How must these new (or repurposed) system functions be designed to safely interact with other system functions?

\paragraph{Persistence} Technologies for ensuring data persistence have evolved and driven changes in computer system architecture.  Until recently, Random Access Memory (RAM) has required a constant source of power to maintain a representation of stored data.  When data stored in RAM is ephemeral, programmers grouped data items together into a file that could be written to secondary storage media (e.g., magnetic media such as drives or tapes).  Recent technology advances are enabling architectural changes at the level of a computer  \cite{2019-Solihin-Persistence} and at the level of a distributed, cloud-hosted file system \cite{2015-Ousterhout-RAMCloud}.  Computer architecture may change considerably due to the development of non-volatile main memory (NVMM) technology, which provides persistent storage (i.e., data remains in the memory after the system is powered off and can be accessed again when it is powered on again), and distributed systems architectures have changed considerably due to the availability of Internet-accessible cloud-hosted storage systems that can store an immense amount of data.  With such technologies in wide use, the meaning and expectation of data persistence is shifting.  Some detectors may be intentionally architected through system design processes.  Is persistence the dual of forgetting?  It is possible that a better understanding of both forgetting and persistence could emerge if these concepts were studied together? Do stronger persistence capabilities necessarily lead to weaker forgetting capabilities (i.e., are the two concepts competitive or synergistic  duals?)

\paragraph{Tools}  Conceptual, software, and hardware tools are all essential to support specification, reasoning, and implementation of forgetting. Without purpose-built forgetting tools at every layer of abstraction, ad hoc attempts to erase information will lead to ongoing misunderstandings about the quality of erasures and additional rounds of cat-and-mouse with cybersecurity researchers and attackers.  How can tools that support forgetting provide high assurance against various detector models and be efficient enough for real-world use?  Performance issues have limited the applicability of system instrumentation techniques for a variety of useful properties.  Can the performance of forgetting tools be designed to scale with the demands of detectors (i.e., a stronger forgetting capability requires more tolerance for  system performance impacts)?

\paragraph{Methodology for effective forgetting} What are best practices for forgetting information?  When to forget?  How should the need for forgetting be balanced against the need for accountability or timeliness in a system?  For example, when an application is designed to assiduously log every action, how should a \forget instruction be interpreted? Over time, software engineers have defined design patterns to capture engineering best practices.  Will these patterns adapt to incorporate forgetting?  Will new patterns emerge as system designers identify best practices for cleaning up or shutting down  that incorporate  forgetting as a step in a larger process?

\paragraph{Forgetting in nature}
Natural processes have been used in support of computing.  For example, natural processes that exhibit randomness, such as cosmic ray flux,  can be used as sources of entropy for strong cryptographic systems.  Are there natural forgetting processes that might provide insights or help to support  strong forgetting mechanisms or measures?  During sleep, synaptic
connections are broken and cleared biochemically to make room for new connections.  Wind and rain erase footprints and tire tracks.   Is there a physics of forgetting that we can leverage?

\paragraph{Artificial intelligence and cloud storage} When a system architect, designer, or developer is building on a well-defined implementation layer, forgetting is challenging even in the presence of weak detectors.  Forgetting will be  more challenging when detectors gain access  to a tremendous volume of data and capabilities for information inference.  For example, an attacker that can access files of passwords previously leaked from a variety of systems may be able to supplement local information to more quickly infer sensitive system information that would be effectively forgotten with respect to locally constrained detectors.  Must the detector concept itself be extended to account for detectors that can use externally gained contextual information to infer local information?   How should we approach intentional forgetting when  powerful AI systems and enormous cloud-based storage systems offer such tremendous reasoning assistance and memory aids to detectors?

To advance the goal of enabling intentional forgetting, research is needed in many domains. The  topics and questions above are only a few examples of research areas that require attention.

\section{Conclusion}
\label{conclusion}

This position paper introduced the concept of intentional forgetting and discussed the capabilities that are needed to enable it.  Intentional forgetting is important for traditional types of applications and systems and emerging IoT and cloud-based systems. Big data and powerful learning and reasoning algorithms amplify the challenges in forgetting by providing powerful observation capabilities.

While  current technologies instrument systems to keep track of data or enable data deletion (e.g., garbage collection, encrypted memory), they do not enable a general forgetting capability. However, these instrumentation technologies offer a foundation of knowledge that can  inform advancements in systems design to support intentional forgetting.  Essential to intentional forgetting is the ability to clearly specify and rigorously reason about  the forgetting capability provided by a system. We have discussed  approaches to modeling computations  at both the system abstraction and  implementation layers, and highlighted the need to model detectors.   

Developing intentional forgetting capabilities that are robust enough to support the cybersecurity and privacy needs of future systems will require research advances in a variety of theoretical and systems-oriented domains.  We reviewed a
number of well developed techniques to build on and use for inspiration. However, real progress will require cross-disciplinary collaboration and fundamentally novel approaches.

We hope readers are intrigued by the challenges and are motivated to contribute to the research  that will one day result in a robust theory and practice of intentional forgetting.

\subsubsection*{Acknowledgements} This work was funded by the U.S. Department of Homeland Security (DHS) Science and Technology (S\&T) Directorate under Contract No. HSHQDC-16-C-00034. Any opinions, findings, and conclusions or recommendations expressed in this material are those of the authors and do not necessarily reflect the views of DHS and should not be interpreted as necessarily representing the official policies or endorsements, either expressed or implied, of DHS or the U.S. Government.

{\footnotesize \bibliographystyle{acm}
\bibliography{forgetting-bib}}


\end{document}